\documentclass[11pt]{article}
\usepackage{graphics}
\usepackage{epsfig}
\usepackage{epstopdf}
\usepackage{amsmath}
\usepackage{array}% bold math
\begin{document}
\begin{center}
{\LARGE Emergence of cosmic space and the maximization of horizon entropy\\[0.2in]}
{ P. B.  Krishna and Titus K. Mathew\\
e-mail:krishnapb@cusat.ac.in, titus@cusat.ac.in \\ Department of
Physics, Cochin University of Science and Technology, Kochi, India.}
\end{center}

\abstract{The spatial expansion of the universe can be described as the emergence of space with the progress of cosmic time, through a
simple equation, $\Delta V = \Delta t\left(N_{surf}- N_{bulk}\right)$. This law of emergence suggested by Padmanabhan in the
context of general relativity for a flat universe has been generalized by Sheykhi to Gauss Bonnet and Lovelock gravity for a universe with any spacial curvature. We 
investigate whether this generalized holographic equipartition effectively implies the maximization of horizon entropy. First, we obtain the constraints
imposed by the maximization of horizon entropy in Einstein, Gauss Bonnet and Lovelock gravities for a universe with any spacial curvature. We then analyse
the consistency of the law of emergence in \cite{Sheykhi}, with these constraints obtained. Interestingly, both the law of emergence and the horizon entropy maximization
demands an asymptotically de Sitter universe with $\omega \geq -1$. More specifically, when the degrees of freedom in the 
bulk
$( N_{bulk})$ becomes equal to the degrees of freedom on the boundary surface $(N_{surf}),$ the universe attains a state of maximum horizon entropy. Thus, the law of
emergence can be viewed as a tendency for maximizing the horizon entropy, even in a non flat universe. Our results points at the deep connection between 
the law of emergence and horizon thermodynamics, beyond Einstein gravity irrespective of the spacial curvature.}

\section{Introduction}

The intriguing connection between gravitational dynamics  and thermodynamics motivates the emergent interpretation of gravity. This curious connection
was first realized 
by Bekenstien and Hawking with the discovery of black hole mechanics \cite{Bekenstein1,Bekenstein2,Hawking1,Hawking2}. A major step along this line was 
taken by Jacobson, who derived Einstein's field equations from the fundamental Clausius relation on a local Rindler causal
 horizon \cite{Jacob1}. Later, Padmanabhan derived Newton's law of gravity from the thermodynamic relations \cite{Paddy2}. In 2011, Verlinde described gravity as an emergent phenomenon by considering it as an entropic force caused due to the changes in the positions
 of the material bodies \cite{verlinde}. Following this idea, several schemes for relating gravity and thermodynamics have been put forward for a variety of gravity theories
 \cite{Eling1,Paddy1}. But, most of these approaches considers the gravitational field as an emergent phenomenon leaving the spacetime
 background as pre-existing.

Recently, a novel perspective for describing gravity was suggested by Padmanabhan, assuming the spacetime itself as an emergent structure \cite{Paddy3}.
Conceptually,
it is hard to imagine time as being emerged from some pregeometric variables. It is also difficult to treat the space around finite gravitating systems 
as emergent. But, these conceptual complexities will disappear in the cosmological context, when one chooses the time variable as the proper time of the 
geodesic observer to whom CMBR appears homogeneous and isotropic. Thus, the emergence of spacetime provide an elegant way of describing the cosmological
evolution as the emergence of cosmic space with the progress of cosmic time. He could also arrive at the Friedmann equation for a spatially flat universe
from this new idea, in the context of Einstein gravity. With appropriate modifications in the surface degrees of freedom, Cai extended this procedure to
higher dimensional gravity theories including Gauss Bonnet and Lovelock gravities for a spatially flat universe \cite{Cai1}. Inspired by this another
generalization to
Padmanabhan's proposal was suggested in \cite{Yang}, where the authors assume a more complex relation to describe the rate of emergence. Following 
\cite{Cai1}, a successful generalization of Padmanabhan's idea to a non flat universe was done by Sheykhi \cite{Sheykhi}. Slightly modifying Cai's proposal,
he could arrive at the Friedmann equations in Einstein, Gauss Bonnet and Lovelock gravities for a universe with any spacial curvature. Eune and Kim proposed
another generalization to Padmanabhan's idea employing the proper invariant volume for describing the rate of emergence in a non flat universe \cite{Eune}. 
But, the authors of \cite{Eune}, had to redefine the Planck length, which is a fundamental constant as a function of cosmic time to deduce the Friedmann 
equations. This idea of emergence have also been extended to the braneworld scenarios \cite{Sheykhi1,Sheykhi2,Sheykhi3}. For more investigations on this idea, see 
\cite{sumanta1,sumanta2,sumanta3,Komatsu,Zhang,Hashemi,MKT,HAKT,KT3,Komatsu1}
 
On the other hand, it is well known that, every ordinary macroscopic system evolves to an equilibrium state of maximum entropy \cite{Callen1}. In \cite{Diego1},
D. Pavon 
and N. Radicella have shown that, our universe with a Hubble expansion history behaves as an ordinary macroscopic system that proceeds to a state of 
maximum entropy. Hence it is natural to think whether the law of emergence proposed by Padmanabhan ensures entropy maximization. In a previous work, we
have proved that the holographic equipartition law in \cite{Paddy3}, proposed in the context of general relativity effectively implies the maximization
of entropy \cite{KT}. Further in \cite{KT2}, we have proved the consistency of the generalized holographic equipartition in \cite{Cai1,Yang} with the 
constraints
of entropy maximization in the context of Einstein, Gauss Bonnet and Lovelock gravity. What was remarkable in our result was that, both the generalizations
in \cite{Cai1} and \cite{Yang} imply the maximization of horizon entropy, even if their basic assumptions are different. But it has to be mentioned that,
the results in \cite{KT} and \cite{KT2} are restricted to a spatially flat universe. Here our aim is to extend the results in \cite{Diego1} to a non flat universe
in the context of Einstein, Gauss Bonnet and Lovelock gravity and to investigate the consistency of the generalized holographic equipartition with the 
constraints of horizon entropy maximization.
 
 We have already mentioned about the two generalizations of Padmanabhan's proposal in a non flat universe \cite{Sheykhi,Eune}. In \cite{Sheykhi}, Sheykhi
 considered the areal volume for describing the rate of emergence and the authors of \cite{Eune} employed proper invariant volume in formulating the law of
 emergence. It is to be noted that, while Sheykhi could easily arrive at the Friedmann equations, the authors of \cite{Eune} have to redefine the Planck length
 as a function of cosmic time in order to reach at the Friedmann equations. The modification of a fundamental constant itself is troublesome. Moreover, the 
 modified Planck length seem to diverge in the matter and radiation dominated era. These problems faced by the choice of invariant volume in formulating
 the law of emergence are also addressed in \cite{HKT}, which supports the use of areal volume in a non flat universe from a thermodynamic point of view.
 It is noteworthy that the authors of \cite{CAI1,CAI2,CAI3} have also used the areal volume in establishing the connection between the Friedmann equations
 and the first law of thermodynamics in different perspectives.
 Due to the above reasons we choose the generalization in \cite{Sheykhi} for the present study. The scope of this paper is to analyze whether the law of
 emergence in \cite{Sheykhi} leads to the maximization of horizon entropy.
 
 The rest of the paper is organized as follows. In section 2, we obtain the constraints for the maximization of horizon entropy in the context of Einstein,
 Gauss Bonnet and Lovelock gravity for a universe with any spacial curvature. In section 3, we will check the consistency of the law of emergence in 
 \cite{Sheykhi} with the constraints of horizon entropy maximization. We present our conclusions in section 4.

\section{Horizon entropy maximization in a non flat universe}
In the context of Einstein gravity, it was shown in \cite{Diego1} that our universe with a Hubble expansion history evolves to an equilibrium state of
maximum entropy consistent with the constraints,
\begin{equation}\label{eqn:conditions1}
 \dot S \geq 0, \, \textrm{always}  
\end{equation} and
\begin{equation}\label{eqn:conditions2}
   \, \, \, \, \, \,  \ddot S <0 \, \, \, \textrm{at least in long run}.
\end{equation}
Here $S$ represents the total entropy of the universe which can be approximated as the horizon entropy and dots denote the derivatives with respect to a
relevant variable. In a previous work, we have extended this procedure to Gauss Bonnet and Lovelock gravity for a spatially flat universe \cite{KT2}. 
Our aim here is
to generalize these results to a universe with a non zero spacial curvature. More specifically, we analyse whether the horizon entropy is getting extremized
in a non flat universe in the framework of Einstein, Gauss Bonnet and Lovelock gravities.

Let us consider an FLRW universe in $(n+1)$ dimensional spacetime with an expansion factor $a(t)$ and spacial curvature $k$, described by the Friedmann
equation,
\begin{equation}\label{eqn:FR1}
 H^2 +\frac{k}{a^2} =\frac{8\pi L^2_p}{3}\rho.
\end{equation}
Here, $H= \frac{\dot a(t)}{a(t)}$ is the Hubble parameter, $\rho$ is the energy density and $L_p$ represents the Planck length.
We can consider any closed surface as the thermodynamic boundary of a system (real or notional), through which energy and matter may enter or leave the system like the event horizon or the apparent horizon. Here we consider the apparent horizon with radius,
\begin{equation}\label{eqn:RA}
 \tilde{ r}_A =\frac{1}{\sqrt{H^2 +\frac{k}{a^2}}},
\end{equation}
as the thermodynamic boundary since the event horizon usually seem to violate the laws of thermodynamics. Motivated by the Bekenstein-Hawking result \cite{Bekenstein1,Hawking1}, the entropy associated with the apparent horizon can be expressed
as,
\begin{equation}\label{eqn:ES}
 S = \frac{A_{n+1}}{4L^{n-1}_p},
\end{equation}
where ${A_{n+1}}=n\Omega_n {\tilde{r}_A}^{n-1}$, for $n\geq 3$, is the area of the apparent horizon with $\Omega_n$ being the volume of the unit $n$ sphere.
 
 We shall now check whether the horizon entropy is getting maximized in course of time.  When the universe reaches its final stage, scale factor will increase and the radius of the apparent horizon asymptotically approaches the Hubble radius. Hence one could guess that the entropy of a non flat universe will approach a maximum finite value in the long run, as in \cite{Diego1}. We can prove it explicitly, with the help of Friedmann equation. Taking the derivative of equation (\ref{eqn:ES}) with respect to the 
 cosmic time, we get,
 
\begin{equation}\label{eqn:dns}
 \dot S = \frac{n(n-1)\Omega_n }{4L^{n-1}_p} \tilde {r}_A^{n-2}\dot{\tilde{r}}_A.
\end{equation}
For satisfying the constraint $\dot S\geq 0$, $\dot{\tilde{r}}_A$ in the above equation should always be greater than or equal to zero.
So the entropy will never decrease, if $\dot{\tilde{r}}_A\geq 0$.
Using equations (\ref{eqn:FR1}) and (\ref{eqn:RA}) along with continuity equation, $\dot\rho+nH(\rho+P)=0$, one can then arrive at,
\begin{equation}\label{eqn:radot}
\dot{ \tilde{r}}_A ={\frac{n}{2} H {\tilde{r}_A}(1+\omega)},
\end{equation}
where $\omega$ is the equation of state parameter defined through the equation of state $P=\omega\rho$. For an asymptotically de Sitter universe with $\omega \geq -1,$ $ \dot{ \tilde{r}}_A$ will be non negative which in turn implies the non negativity of $\dot S$.

Now, we will check whether this entropy attains a maximum value in the long run characterized by the inequality  $\ddot S< 0$. Taking the time derivative
of equation (\ref{eqn:dns}), we obtain
\begin{equation}\label{eqn:ddns}
 \ddot S = \frac{n(n-1)\Omega_n }{4L^{n-1}_p} \tilde {r}_A^{n-3}[{(n-2)\dot{ \tilde{r}}_A^2}+{\tilde{r}_A}{\ddot{\tilde{r}}_A}].
\end{equation}
In order to satisfy the constraint, $\ddot S< 0$,  ${\ddot{\tilde{r}}_A}$ in the above equation should be less than zero in the asymptotic limit. 
The maximization
of entropy also demands,
\begin{equation}\label{eqn:ccon2}
  |(n-2){\dot{ \tilde{r}}_A^2}|<|{\tilde{r}_A}{\ddot{\tilde{r}}_A}|,
\end{equation}
to be satisfied at least in the last stage of evolution.
To verify these, we first take the time derivative of equation (\ref{eqn:radot}), 
\begin{equation}\label{eqn:raddot}
\ddot{ \tilde{r}}_A ={\frac{n}{2} { {\tilde{r}_A}[\frac{n}{2}(1+\omega)^2 H^2+(1+\omega)\dot H+\dot \omega H}]}.
\end{equation}
In the final de Sitter epoch, $\omega \to -1$ and all terms containing $(1+\omega)$ vanishes. Since $\dot \omega$ is always negative, the above equation
guarantees the condition
$\ddot{ \tilde{r}}_A<0$, in the long run. Also, in the asymptotic limit, $t\to \infty$, $\dot{ \tilde{r}}_A $ will tend to zero as $\omega \to -1$ .
Then the inequality in (\ref{eqn:ccon2}) readily follows ensuring maximization of entropy. Thus, in the context of Einstein's gravity, the horizon entropy
of a non flat universe will never grow unbounded. 

It has to be mentioned that most of the quintessence-like dark energy models predicts a universe, which evolves to a pure de Sitter state with $\omega\geq -1$. The results in \cite{Mimoso} , also demands a cosmological constant dominated de sitter expansion for the universe, to behave as an ordinary thermodynamic system, excluding the domination of phantom fields. Even though, there are phantom dark energy models with $\omega< -1$, such models usually suffer from quantum instabilities and thus can not be considered as useful dark energy models \cite{Carrol,Cline}.

Now, we will extend this procedure to Gauss Bonnet gravity, which is a natural extension of Einstein's gravity such that the Gauss Bonnet action contains
second order terms too. The entropy relation for the black hole horizon in Gauss Bonnet gravity is assumed to be satisfied for the FLRW universe also. 
Following this, the entropy of the apparent horizon takes more complex form unlike in the Einstein's gravity, which can be expressed as,
\cite{Cai2,Cai3}, 
\begin{equation}\label{eqn:gbs}
 S=\frac{A}{4 L^{n-1}_p}\left[1+\frac{n-1}{n-3}\frac{2\tilde\alpha}{\tilde{r}_A^{2}}\right]
\end{equation}
where $A=n\Omega_n {\tilde{r}_A}^{n-1}$, for $n\geq4$, is the area of the apparent horizon and $\tilde\alpha =(n-2)(n-3)\alpha$, with $\alpha$ being the Gauss Bonnet 
coefficient which is positive \cite{DGB}.
From the above relation, we can immediately obtain the rate of change of horizon entropy as,
\begin{equation}\label{eqn:gbdns}
 \dot S = \frac{n(n-1)\Omega_n }{4L^{n-1}_p} \tilde {r}_A^{n-2}\dot{\tilde{r}}_A (1+{2\tilde\alpha \tilde{r}_A^{-2}}).
\end{equation}
Now, the Friedmann equation in Gauss Bonnet gravity can be written as,
\begin{equation}
 \tilde{r}_A^{-2} + {\tilde\alpha \tilde{r}_A^{-4}}=\frac{16\pi L^{n-1}_p}{n(n-1)}\rho
\end{equation}
which on differentiation gives
\begin{equation}\label{eqn:gbdra}
\dot{ \tilde{r}}_A =\frac{8\pi L^{n-1}_p}{(n-1)}{{\tilde{r}_A^{3}H} { (1+\omega)\rho}\over (1+{2\tilde\alpha \tilde{r}_A^{-2}})}.
\end{equation}
Since $\omega$ is always greater than or equal to $-1$, for an asymptotically de Sitter universe, the above equation ensures the condition $\dot S\geq 0$.

In order to obtain the constraint for the entropy maximization, we find the second derivative of horizon entropy as,
\begin{equation}\label{eqn:gbddns}
 \ddot S = \frac{n(n-1)\Omega_n }{4L^{n-1}_p} \tilde {r}_A^{n-3}\left[{\dot{\tilde{r}}_A}^{2}[{(n-2)+(2n-8)\tilde\alpha{ \tilde{r}}_A^{-2}}]+[{\tilde{r}_A}{\ddot{\tilde{r}}_A}(1+{2\tilde\alpha \tilde{r}_A^{-2})}]\right].
\end{equation}
For the non positivity of $\ddot S $, ${\ddot{\tilde{r}}_A}$ should be less than zero in the long run. Moreover, from the above equation, we immediately
reach at the constraint for the horizon entropy maximization,
\begin{equation}\label{eqn:con3}
  |{\dot{ \tilde{r}}_A^2}[{(n-2)+(2n-8)\tilde\alpha{ \tilde{r}}_A^{-2}}]|<|{\tilde{r}_A}{\ddot{\tilde{r}}_A}(1+{2\tilde\alpha \tilde{r}_A^{-2}})|.
\end{equation}
Differentiating equation
(\ref{eqn:gbdra}), we find
\begin{equation}\label{eqn:gbraddot}
\ddot{ \tilde{r}}_A =\frac{8\pi L^{n-1}_p}{(n-1)} {{\tilde{r}_A^{3}} { \rho}\over (1+{2\tilde\alpha \tilde{r}_A^{-2}})} \left [{3\over 2}(1+\omega)^2 H^2+(1+\omega)\dot H+\dot \omega H+{4\tilde\alpha H
{ (1+\omega){\tilde{r}_A^{-3}} \dot{\tilde{r}}_A}\over (1+{2\tilde\alpha \tilde{r}_A^{-2}})}\right].
\end{equation}
Since $\omega$ approaches $-1$ in the final stage, all terms containing $(1+\omega)$ vanishes. Then, ${\ddot{\tilde{r}}_A}$ will be negative in the final
stage as the state parameter '$\omega$' is a decreasing function. Also, Since ${\dot{ \tilde{r}}_A^2}\to 0$, when $t\to \infty$, the constraint in
(\ref{eqn:con3}) will hold true in the final stage. Thus, the horizon entropy of a non flat universe is getting maximized in the final stage, in the context
of Gauss Bonnet gravity.

We will now move to the Lovelock gravity, which is a generalization of Gauss Bonnet gravity such that the Lagrangian consists of dimensionally extended
Euler densities \cite{Love1}. The entropy relation for black holes in Lovelock gravity is assumed to be hold for the apparent horizon of the FLRW universe which can be
defined as,
\begin{equation}\label{eqn:lls}
 S= \frac{A}{4L^{n-1}_p} \sum_{i=1}^m \frac{i(n-1)}{(n-2i+1)} \hat{c_i}{\tilde r_A}^{2-2i}
\end{equation}
 where $\, \, \, \displaystyle m=n/2$ and the coefficients $\, \, \, \, \hat c_i$ are given by $\displaystyle \, \, \, \, \, \hat c_0=\frac{c_0}{n(n-1)} , 
 \hat c_1 =1,  \hat c_i = c_i \prod_{j=3}^{2m}(n+1-j) \, i>1$.
Differentiating the above equation, we get
\begin{equation}\label{eqn:llds}
 \dot S= \frac{n\Omega_n {\tilde{r}_A}^{n}{\dot{\tilde{r}}_A}}{4L^{n-1}_p} \sum_{i=1}^m {i(n-1)} \hat{c_i}{\tilde r_A}^{-2i}.
\end{equation}
Here the change in entropy, $\dot S$ will always be non negative if $\dot{\tilde{r}}_A$ is always greater than or equal to zero.
Now, using the Friedmann equation in Lovelock gravity given by,
\begin{equation}
 \sum_{i=1}^m  \hat{c_i}{\tilde r_A}^{-2i} =\frac{16\pi L^{n-1}_p}{n(n-1)}\rho,
\end{equation}
one can arrive at
\begin{equation}\label{eqn:lldra}
 {\dot{\tilde{r}}_A} =\frac{16\pi L^{n-1}_pH(1+\omega)\rho}{\sum_{i=1}^m{2i}  \hat{c_i}{\tilde r_A}^{-2i-1}}.
\end{equation}
Since $\omega\geq-1$, for an asymptotically de Sitter universe, the above equation guarantees the non negativity of ${\dot{\tilde{r}}_A}$ which in turn
ensures the non negativity of $\dot S$ as per equation (\ref{eqn:llds}).

We will now obtain the constraint for the maximization of horizon entropy. Differentiating equation (\ref{eqn:llds}), we could easily find,
\begin{equation}\label{eqn:lldds}
\ddot S= \frac{n(n-1)\Omega_n {\tilde{r}_A}^{n-1}}{4L^{n-1}_p} \sum_{i=1}^m i \hat{c_i}{\tilde r_A}^{-2i}[{(n-2i)\dot{ \tilde{r}}_A^2}+{\tilde{r}_A}{\ddot{\tilde{r}}_A}].
\end{equation}
As the first term in the above equation is always positive, the condition $\ddot S<0$ requires the non positivity of ${\ddot{\tilde{r}}_A}$ in the long run.
Differentiating, equation (\ref{eqn:lldra}), one readily gets,
\begin{equation}\label{eqn:llraddot}
\ddot{ \tilde{r}}_A =\frac{16\pi L^{n-1}_p \rho}{(n-1)\sum_{i=1}^m{2i}  \hat{c_i}{\tilde r_A}^{-2i-1}  }  \left [(1+\omega)\dot H+\dot \omega H-n(1+\omega)^2 H^2-
(1+\omega)H\frac{\dot{\tilde{r}}_A}{{\tilde{r}}_A} \right].
\end{equation}
Since $\omega\to -1$, when $t\to\infty$, all terms containing $(1+\omega)$ vanishes. Then, as $\dot\omega$ being negative, the above equation guarantees
the non positivity of $\ddot{ \tilde{r}}_A$. The non positivity of $\ddot S$ also demands,
\begin{equation}\label{eqn:con4}
  |\sum_{i=1}^m  i\hat{c_i}{\tilde r_A}^{-2i}(n-2i){\dot{ \tilde{r}}_A^2}|<|\sum_{i=1}^m  i\hat{c_i}{\tilde r_A}^{-2i}{\tilde{r}_A}{\ddot{\tilde{r}}_A}|
\end{equation}
to be satisfied in the asymptotic limit as per equation (\ref{eqn:lldds}).
As ${\dot{\tilde{r}}_A}\to 0$, in the final stage, the above inequality holds true indicating the maximization of horizon entropy.

Here, we have obtained the constraints for the maximization of horizon entropy in the context of Einstein, Gauss Bonnet and Lovelock
gravities and have seen that these constraints are satisfied by an asymptotically de Sitter universe with $\omega\geq -1$ in all
these gravity theories irrespective of the spacial curvature.

\section{The law of emergence and the horizon entropy maximization}
This section starts with a brief review of Padmanabhan's proposal in \cite{Paddy3}, followed by a discussion on the the formulation of the law of emergence
in \cite{Sheykhi}, on which our work is closely related. Then, we will see whether the law of emergence implies horizon entropy maximization in a non flat
universe in the context of Einstein, Gauss Bonnet and Lovelock gravities. More specifically, we will analyse whether the law of emergence proposed by Sheykhi
in \cite{Sheykhi}, leads to the maximization of horizon entropy.

\subsection{The law of emergence}
Based on the 'Emergent gravity paradigm'\cite{Paddy11,Paddy12}, which was established in the past decades, Padmanabhan interpreted the evolution of the universe as the emergence of cosmic space with the progress of cosmic time. It is observed that a pure de Sitter universe obeys a specific version of the holographic principle in the form,
\begin{equation}
N_{surf} = N_{bulk}.
\end{equation}
Here $N_{surf}$ is the number of degrees of freedom on the surface of the Hubble sphere with Hubble radius, $H^{-1}$ and  $N_{bulk}$ is the number of degrees of freedom residing in the bulk region of space
enclosed by the horizon. One can attribute one degree of freedom per Planck area and thus define the surface degrees of freedom as,
\begin{equation}\label{eqn:Nsurf1}
N_{surf} = \frac{4 \pi}{ L^2_p H^2}.
\end{equation}
Here, we have chosen $'L_p^2=\frac{G\hbar}{c^3}'$ to denote one degree of freedom since $L_p$ act as the lower bound of length scales that can be operationally defined. The bulk degrees of freedom is given by,
\begin{equation}\label{eqn:Nbulk1}
N_{bulk} =\frac{|E|}{\frac{1}{2} {k_B T }},
\end{equation}
where $|E|=|\rho+3p|V$, the Komar energy inside the Hubble volume, $V=\frac{4\pi}{ 3 H^3}$; $k_B$, the Boltzmann constant and $T = \frac{H}{2\pi}$ is
the Gibbon's Hawking temperature. $N_{bulk}$ is a dimensionless number associated with the bulk energy $E_{Komar}$, if the Komar energy inside the bulk volume is at equipartition at the temperature $T$. One should not be confused with the temperature $T$ and $N_{bulk}$ as the normal kinetic temperature of the matter and the standard degrees of freedom. Instead, we have to think that, these degrees of freedom are already emerged, from some pre geometric variables, along with space. When the Hubble radius is endowed with the horizon temperature, it is possible to treat $N_{bulk}$ that have emerged already, along with space, as if they are inside a microwave oven having a temperature which is set to the surface value.

From the definitions of $N_{surf}$ and $ N_{bulk}$, equation (\ref{eqn:Nbulk1}) can be expressed in the form,
$|E|={{1\over2}{k_B T N_{surf}}}$. This equation represents holographic equipartition since it relates the degrees of freedom in the bulk which is determined by the equipartition condition to the degrees of freedom on the horizon surface. If our universe is evolving to a pure de Sitter phase that satisfies holographic equipartition, it may be argued that the evolution of the universe which is equivalent to the emergence of space is due to its tendency for satisfying holographic equipartition. In other words, one can think that the accelerated expansion of the universe is driven by the holographic discrepancy, $'N_{surf} - N_{bulk}'$. The most natural form of such a law will be,
\begin{equation}
\Delta V={\Delta t(N_{surf}-\epsilon N_{bulk})}
\end{equation}
where $V$ is the Hubble volume in Planck units and $t$ is the cosmic time, both in Planck units. This relation can be elevated into a postulate governing the emergence of space by reintroducing Planck length and by setting $\Delta V/\Delta t= dV/dt$, which can be expressed as,
\begin{equation} \label{eqn:dVdt1}
{dV\over dt} ={L^2_P(N_{surf}- \epsilon N_{bulk})}.
\end{equation} One can easily show that, the above expression simplifies to the Friedman equation. Thus, using the concept of holographic equipartition, the evolution of the universe can be described as the emergence of cosmic space with the progress of cosmic time.

The first successful extension of Padmanabhan's conjecture to a non flat universe was done by Sheykhi \cite{Sheykhi}. In this proposal the rate of volume 
increase is assumed to be proportional to the difference in the degrees of freedom, $(N_{surf}-  N_{bulk})$ (as in Padmanabhan's original proposal), but,
the function of proportionality is assumed as the ratio of the apparent horizon radius and the Hubble radius. Thus, the law of emergence in $(n+1)$ 
dimensional space will take the form,
\begin{equation} \label{eqn:dVdt3}
 \alpha\frac{dV}{dt} ={L^{n-1}_p\frac{\tilde{r}_A}{H^{-1}}(N_{surf}- N_{bulk})}.
\end{equation}
where, $V=\Omega_n {\tilde{r}_A}^{n}$, with $\Omega_n$ being the volume of the unit n-sphere.
For a non flat universe in $(n+1)$ dimensional Einstein's gravity, the degrees of freedom on the apparent horizon can be defined as,
\begin{equation}\label{eqn:Nsurf3}
 N_{surf} =\frac{ \alpha A}{ L^{n-1}_p}
\end{equation}
where $A=n\Omega_n {\tilde{r}_A}^{n-1}$ and $\alpha =\frac{n-1}{2(n-2)}$.
Assuming the temperature associated with the apparent horizon, $T$ as $\frac{1}{2\pi \tilde{r}_A}$, the bulk degrees of freedom can be defined as,
\begin{equation}\label{eqn:Nbulk3}
 N_{bulk} =-4\pi\Omega_n{\tilde{r}_A}^{n+1} \frac{(n-2)\rho +np }{n-2}.
\end{equation}
Substituting (\ref{eqn:Nsurf3}) and (\ref{eqn:Nbulk3}) in equation(\ref{eqn:dVdt3}), one could easily arrive at the Friedmann equation of an $(n+1)$ 
dimensional FLRW universe with any spacial curvature.

We will now turn our attention to Gauss Bonnet gravity. From the definition of entropy in (\ref{eqn:gbs}), the  increase in effective volume in Gauss 
Bonnet gravity could be calculated as,
\begin{equation}\label{eqn:Veff1}
 \frac{d\tilde V}{dt} =n\Omega_n \dot{\tilde r}_A {\tilde{r}_A}^{n-1}(1+2\tilde\alpha \tilde{r}_A^{-2})
\end{equation}
Using the above relation,  Sheykhi defined the surface degrees of freedom as,
\begin{equation}\label{eqn:Nsurf4}
  N_{surf}=\frac{\alpha n\Omega_n {\tilde{r}_A}^{n-1}}{L^{n-1}_p}(\tilde{r}_A^{-2}+\tilde\alpha \tilde{r}_A^{-4}).
\end{equation}
The degrees of freedom residing in the bulk is still given by (\ref{eqn:Nbulk3}). Employing (\ref{eqn:Nbulk3}),(\ref{eqn:Veff1}),(\ref{eqn:Nsurf4}) in
equation
(\ref{eqn:dVdt3}); and by replacing $V\to \tilde V, $ one could arrive at the Friedmann equation in Gauss Bonnet gravity for a universe with any spacial
curvature.

Now, we will move to the Lovelock gravity. From equation (\ref{eqn:lls}), one could calculate the volume increase in Lovelock gravity as,
\begin{equation}\label{eqn:Veff2}
 \frac{d\tilde V}{dt}= n\Omega_n {\tilde{r}_A}^{n+1} 
 \left(\sum_{i=1}^m i \hat{c_i}{\tilde r_A}^{-2i}\right)\dot{\tilde r}_A.
\end{equation} 
From the above relation, the surface degrees of freedom could be defined as,
\begin{equation}\label{eqn:Nsurf5}
 N_{surf}= \frac{\alpha n\Omega_n }{L^{n-1}_p} {\tilde{r}_A}^{n+1} 
 \sum_{i=1}^m  \hat{c_i}{\tilde r_A}^{-2i}.
\end{equation}
From equations (\ref{eqn:Nbulk3}), (\ref{eqn:Veff2}), (\ref{eqn:Nsurf5}) and (\ref{eqn:dVdt3}) one could reach at the Friedmann equation in Lovelock 
gravity for a universe with any spacial curvature.

\subsection{Holographic equipartition and the horizon entropy maximization}
Now, we will check the consistency of the generalized holographic equipartition in \cite{Sheykhi} with the constraints of horizon entropy maximization.
Consider an $(n+1)$ dimensional FLRW universe in Einstein's gravity with a spacial curvature $k$. The rate of change of cosmic volume,
$V=\Omega_n {\tilde{r}_A}^{n}$, with respect to the cosmic time could be obtained as,
\begin{equation}
  {dV\over dt}=n\Omega_n {\tilde{r}_A}^{n-1}{\dot{\tilde{r}}_A}.
\end{equation}
Comparing the above relation with equation (\ref{eqn:dns}), we can express the rate of emergence as,
\begin{equation}\label{eqncon1}
 \dot S ={\frac{(n-2)H}{2}(N_{surf}- N_{bulk})}.
\end{equation}
For the non negativity of $\dot S$, the holographic discrepancy, $'{N_{surf}- N_{bulk}}'$ in the above equation should be greater than or equal to zero.
We will now see whether the definitions of ${N_{surf}}$ and $ {N_{bulk}}$ ensures the non negativity of $\dot S$.
Recalling eq.(\ref{eqn:Nsurf3}), we have,
\begin{equation}\label{eqn:Nsurf31}
 N_{surf} = {\frac{n(n-1)}{2(n-2)}}  \frac{\Omega_n {\tilde{r}_A}^{n-1}}{ L^{n-1}_p}.
\end{equation}
Using the Friedmann equation in Gauss Bonnet gravity, the bulk degrees of freedom in (\ref{eqn:Nbulk3}) can be expressed as,
\begin{equation}\label{eqn:Nbulk31}
 N_{bulk} = {\frac{n(n-1)}{2(n-2)}}  \frac{\Omega_n {\tilde{r}_A}^{n-1}}{ L^{n-1}_p}-
 {\frac{n(n-1)}{2(n-2)}}  \frac{\Omega_n {\tilde{r}_A}^{n-2}{\dot{\tilde{r}}_A}}{ L^{n-1}_p H}.
\end{equation}
From the above equations, one can express the holographic discrepancy as,
\begin{equation}
N_{surf}-N_{bulk} = {\frac{n(n-1)}{2(n-2)}}  \frac{\Omega_n {\tilde{r}_A}^{n-2}{\dot{\tilde{r}}_A}}{ L^{n-1}_p H}.
\end{equation}
In order to ensure satisfy the condition $'{N_{surf}- N_{bulk}}\geq 0'$, ${\dot{\tilde{r}}_A}$ in the above expression should always be non negative.
If the universe is assumed to be asymptotically de Sitter, ${\dot{\tilde{r}}_A}$ will be greater than or equal to zero ensuring the non negativity 
of $\dot S$.

Now, we will see whether this horizon entropy is getting maximized in the long run. For that, we obtain the second derivative of horizon entropy from
eq.(\ref{eqncon1}) as,
\begin{equation}\label{eqncon11}
 \ddot S ={\frac{(n-2)\dot H}{2}(N_{surf}- N_{bulk})}+{{\frac{(n-2)H}{2}} {d\over dt}(N_{surf}- N_{bulk})}. 
\end{equation}
In the final de Sitter stage, $N_{bulk}$ approaches $N_{surf}$ and the first term in the above expression vanishes. Since $ {N_{bulk}}$ does not exceed 
$N_{surf}$, $'{N_{surf}- N_{bulk}}'$, will always  be positive and tending to zero in the final stage. In consequence we have,
\begin{equation}
 {d\over dt}(N_{surf}- N_{bulk})<0
\end{equation}
ensuring the non positivity of $\ddot S$. In other words, our universe is trying to minimize the holographic discrepancy and thus guarantees the non 
positivity of $\ddot S$ in the long run. Also, combining equations (\ref{eqn:Nsurf31}), (\ref{eqn:Nbulk31}) and (\ref{eqncon11}), one could easily obtain
the condition for the non positivity of  $\ddot S$ as,
\begin{equation}\label{eqn:con2}
  |(n-2){\dot{ \tilde{r}}_A^2}|<|{\tilde{r}_A}{\ddot{\tilde{r}}_A}|,
\end{equation}
which is same as the constraint in (\ref{eqn:ccon2}), that we have obtained for the maximization of horizon entropy in the previous section.
Thus, the law of emergence leads to the maximization of horizon entropy even in a non flat universe in the context of Einstein's gravity.

We shall now see, whether the law of emergence proposed in the context of Gauss Bonnet gravity leads to the maximization of horizon entropy. Combining
(\ref{eqn:Veff1}) and (\ref{eqn:gbdns}), one can relate the rate of emergence to the rate of change of entropy as,
\begin{equation}
 {dV\over dt} = \frac{4L^{n-1}_p }{(n-1)} \tilde {r}_A \dot S
\end{equation}
Then, the law of emergence in eq.(\ref{eqn:dVdt3}) could be expressed as,
\begin{equation}\label{gbeqncon1}
 \dot S ={\frac{(n-2)H}{2}(N_{surf}- N_{bulk})}.
\end{equation}
The holographic discrepancy $'N_{surf}-N_{bulk}'$ in the above equation could be calculated from the relations in 
(\ref{eqn:Nbulk3}) and (\ref{eqn:Nsurf4}) as,
\begin{equation}\label{eqn:hodis}
N_{surf}-N_{bulk} = {\frac{n(n-1)}{2(n-2)}}  \frac{\Omega_n {\tilde{r}_A}^{n-2}{(1+{2\tilde\alpha \tilde{r}_A^{-2}})\dot{\tilde{r}}_A}}{ L^{n-1}_p H}
\end{equation}
For the non negativity of $'\dot S'$, one should have the condition ${\dot{\tilde{r}}_A}\geq0$, 
as per the above equations, which is generally satisfied  by an asymptotically de Sitter universe.

Now we will find the second derivative of horizon entropy by differentiating eq.(\ref{gbeqncon1}) as, 
\begin{equation}\label{eqncon111}
 \ddot S ={\frac{(n-2)\dot H}{2}(N_{surf}- N_{bulk})}+{{\frac{(n-2)H}{2}} {d\over dt}(N_{surf}- N_{bulk})}. 
\end{equation}
In the final stage of evolution $N_{bulk}$ will be equal to $N_{surf}$ and the first term in the above equation vanishes. As the universe is trying to
reduce the holographic discrepancy, we have the condition,
\begin{equation}
 {d\over dt}(N_{surf}- N_{bulk})<0
\end{equation}
which implies the non positivity of $\ddot S $ in the long run. Also, substituting eq.(\ref{eqn:hodis}) in (\ref{eqncon111}), 
one immediately gets the constraint for the non positivity of $\ddot S$ in the long run as,
\begin{equation}\label{eqn:con31}
  |{\dot{ \tilde{r}}_A^2}[{(n-2)+(2n-8)\tilde\alpha{ \tilde{r}}_A^{-2}}]|<|{\tilde{r}_A}{\ddot{\tilde{r}}_A}(1+{2\tilde\alpha \tilde{r}_A^{-2}})|
\end{equation}
which is same as the inequality in (\ref{eqn:con3}), that we have obtained in the previous section as the constraint for the maximization of
horizon entropy. Thus, the law of emergence guarantees the maximization of horizon entropy in the context of Gauss Bonnet gravity in a non flat universe.

We will now generalize this procedure to more general Lovelock gravity. Recalling the equations (\ref{eqn:dVdt3}), (\ref{eqn:Veff2}) and(\ref{eqn:llds}),
the law of emergence 
in Lovelock gravity can be written as,
\begin{equation}\label{lleqncon111}
 \dot S ={\frac{(n-2)H}{2}(N_{surf}- N_{bulk})}
\end{equation}
just as in the previous case. Note that the above relation have the same form as in Einstein and Gauss Bonnet gravities even if $'S'$, $'{N_{surf}}'$ and 
$'{N_{bulk}}'$ have different definitions in accordance with the corresponding gravity theory. Now, recalling equations (\ref{eqn:Nsurf5}) and 
(\ref{eqn:Nbulk3}), the holographic discrepancy can be obtained as,
 \begin{equation}\label{hodis1}
N_{surf}-N_{bulk} = {\frac{n(n-1)}{2(n-2)}}  \frac{\Omega_n {\tilde{r}_A}^{n+1}{\dot{\tilde{r}}_A}}{ L^{n-1}_p H}\sum_{i=1}^m  i\hat{c_i}{\tilde r_A}^{-2i-1}.
\end{equation}
Since, ${\dot{\tilde{r}}_A}\geq0$, for an asymptotically de Sitter universe,  
$ {N_{bulk}}$ will never exceed ${N_{surf}}$, ensuring the non negativity of $\dot S$. Now, differentiating eq.(\ref{lleqncon111}), we get
\begin{equation}\label{lleqncon1111}
 \ddot S ={\frac{(n-2)\dot H}{2}(N_{surf}- N_{bulk})}+{{\frac{(n-2)H}{2}} {d\over dt}(N_{surf}- N_{bulk})}.
\end{equation}
As $ {N_{bulk}}$ approaches ${N_{surf}}$ in the final stage, the first term in the above expression will reduce to zero in the asymptotic limit, $t\to\infty$.
Then, the non positivity of $\ddot S$ demands,
\begin{equation}
 {d\over dt}(N_{surf}- N_{bulk})<0
\end{equation}
in the long run. As the holographic discrepancy is getting reduced with the progress of cosmic time, the above inequality will be satisfied, ensuring the 
non positivity of $\ddot S$. Moreover, using eq.(\ref{hodis1}) and eq.(\ref{lleqncon1111}), we can obtain the constraint for the maximization of horizon 
entropy as,
\begin{equation}\label{eqn:ccon4}
  |\sum_{i=1}^m  i\hat{c_i}{\tilde r_A}^{-2i}(n-2i){\dot{ \tilde{r}}_A^2}|<|\sum_{i=1}^m  i\hat{c_i}{\tilde r_A}^{-2i}{\tilde{r}_A}{\ddot{\tilde{r}}_A}|,
\end{equation}
which is same as the inequality in (\ref{eqn:con4}), that we have obtained as the constraint to be satisfied for the maximization of horizon entropy
in the context of Lovelock gravity. Thus, the law of emergence proposed in \cite{Sheykhi} guarantees the maximization of horizon entropy.

As per the above discussions, one can interpret the law of emergence as a tendency for maximizing the horizon entropy in the context of Einstein, Gauss
Bonnet and Lovelock gravities even in a non flat universe. If our universe obeys the law of emergence, its horizon entropy will never grow unbounded.
In other words, the tendency for satisfying holographic equipartition could be viewed as a tendency for maximizing the horizon entropy for a universe with
any spacial curvature.

\section{Conclusion}
This paper addresses two questions, 1.'whether the horizon entropy of a non flat universe is getting maximized in course of time', 2. 'whether the emergence
of space implies the maximization of horizon entropy in a non flat universe'. In previous works, we have analysed the consistency of the law of emergence
with the constraints of horizon entropy maximization in the context of Einstein, Gauss Bonnet and Lovelock gravities \cite{KT,KT2}. But, it is important to note that
these results restrict to a spatially flat universe. The central theme of this paper is to generalize these results to a non flat universe and to bring out its implications.
We have chosen the modified law of emergence in \cite{Sheykhi}, for the present study, in which the author considered the areal volume for describing the 
rate of emergence of space. The immediate question one may ask is 'why shouldn't we consider the generalization in \cite{Eune}, where the authors employed
the proper invariant volume in formulating the law of emergence'. The major reason is the modification of the fundamental constant, Planck length in 
\cite{Eune}. Besides that, the modified Planck length seem to diverge in the matter and radiation dominated epoch. Moreover, our previous result in 
\cite{HKT}, strongly support the use of areal volume in a non flat universe from a thermodynamic point of view.

It has been proved that, the horizon entropy of the universe tends to a finite maximum, in the context of Einstein gravity \cite{Diego1}. We have extended
this procedure to Gauss Bonnet and more general Lovelock gravity \cite{KT2}. But, it has to be mentioned that these results will hold only in a spatially
flat universe. Our first first aim was to extend these results to a non flat universe. Usually, the thermodynamic quantities in a non flat universe are
related to the apparent horizon, instead of Hubble horizon. Hence, here we assume the apparent horizon as the thermodynamic boundary of the universe. 
First, we obtained the constraints imposed by the maximization of horizon entropy and further investigated the consistency of a non flat universe with 
those constraints. With the help of continuity equation, it is proved that, an asymptotically de Sitter universe with $\omega\geq-1$, evolves to a state
of maximum horizon entropy, in the context of Einstein, Gauss Bonnet and Lovelock gravity.

Our second aim was to analyse the consistency of the law of emergence in \cite{Sheykhi}, with the maximization of horizon entropy in a non flat universe.
From the law of emergence in \cite{Sheykhi}, we have found that the condition $\dot S\geq 0$ implies $'(N_{surf}- N_{bulk})\geq 0'$. Since 
$ N_{bulk}$ does not exceed $N_{surf}$, $\dot S$ will never be negative. Likewise, imposing the constraint $\ddot S<0$, on the law of emergence
leads to the condition ${d\over dt}(N_{surf}- N_{bulk})<0$, irrespective of the gravity theories. As the universe is trying to decrease the holographic
discrepancy, $'N_{surf}- N_{bulk}'$, this inequality will be satisfied in the long run. Moreover, we have shown that these conditions are compatible with
the corresponding constraints of horizon entropy maximization in the respective gravity theories. Thus, the law of emergence ensures the maximization of
horizon entropy even in a non flat universe. Remarkably, both the holographic equipartition and the horizon entropy maximization leads to the same 
constraints, which are generally satisfied by an asymptotically de Sitter universe. It is worth mentioning that, both the law of emergence and the horizon
entropy maximization demands an asymptotically de Sitter universe with $\omega\geq-1$, irrespective of the spacial curvature. In other words, when the
universe is trying to decrease the holographic discrepancy, its horizon entropy is getting maximized. Thus, the emergence of cosmic space can be viewed as
a tendency for maximizing the horizon entropy, even in a non flat universe, in the context of Einstein, Gauss Bonnet and Lovelock gravity.

%{\bf \large Acknowledgement}

%P.B. Krishna acknowledges KSCSTE, Govt. of Kerala for financial support

\end{document}